\documentclass[aps,pre,twocolumn,showpacs,preprintnumbers,superscriptaddress]{revtex4}

\usepackage{graphicx}
\usepackage{epsf} 
\usepackage{dcolumn}
\usepackage{bm}
\usepackage{amssymb}
\usepackage{amsmath}

\begin{document}

\title{Nonequilibrium phase transition in a model for the propagation of
innovations among economic agents}

\author{Mateu Llas}
\author{Pablo M. Gleiser}
\affiliation{Departament de F\'{\i}sica Fonamental, Universitat de
Barcelona, Avda. Diagonal 647, E-08028 Barcelona, Spain}
\author{Juan M. L\'opez}
\affiliation{Instituto de F\'{\i}sica de Cantabria (CSIC--UC),
E-39005 Santander, Spain}
\author{Albert D{\'\i}az-Guilera}
\affiliation{Departament de F\'{\i}sica Fonamental, Universitat de
Barcelona, Avda. Diagonal 647, E-08028 Barcelona, Spain}

\date{\today}

\begin{abstract}
We characterize the different morphological phases that occur in a
simple one-dimensional model of propagation of innovations among
economic agents [X.\ Guardiola, {\it et. al.}, Phys. Rev E {\bf
66}, 026121 (2002)]. We show that the model can be regarded as a
nonequilibrium surface growth model. This allows us to demonstrate
the presence of a continuous roughening transition between a flat 
(system size independent fluctuations) and a rough phase (system size
dependent fluctuations). Finite-size scaling studies at the
transition strongly suggest that the dynamic critical transition
does not belong to directed percolation and, in fact, critical
exponents do not seem to fit in any of the known universality
classes of nonequilibrium phase transitions. Finally, we present
an explanation for the occurrence of the roughening transition and
argue that avalanche driven dynamics is responsible for the novel
critical behavior.

\end{abstract}


\maketitle

\section{Introduction}
In the past few years there has been an increasing interest among
theoretical physicists in complex phenomena occurring in fields
that are far apart from the traditional realm of Physics like
Social and Economic sciences \cite{albert,newman,econo,bouchaud}.
The main reason being that social and economic systems often exhibit many
instances of complex dynamics, including self-organization,
pattern formation, synchronization and phase transition-like
phenomena that closely resemble those observed in nonequilibrium
physical systems \cite{arenas,fernando,castellano,klemm}.
Physicists approach to these systems usually
provides insights into the basic ingredients that should be included
in simple models in order to obtain the dynamics observed.
Although it is clear that Physics inspired models of socioeconomic
phenomena are often very simplistic views of very complicated
systems, the aim is to show how complex macroscopic dynamics might
arise from rather simple rules operating at the 'microscopic' level of
individual agents and their mutual interactions.

In this paper we consider a very simple model of innovation
propagation dynamics in an  economic system formed by
agents \cite{guardiola,Ebeling}.
The aim is to describe in a simple way the adoption of
innovations that occurs among industries, firms or individuals.
Once a brand new product appears in the market, the agents should
decide whether or not they will incorporate the new technology.
Adopting the new technology (in the form of a software, device,
gadget, etc) has a cost, but at the same time it may improve
business performance in the case of firms, or may level off life
quality for individuals.
Innovations are regarded here in a broad sense and stand for any
device or tool. For instance, a firm can decide to incorporate WWW
technology by creating or revamping its WWW page, or going into
e-commerce for the first time. A layman observation is that, if
not always, in most cases, when the new technology actually
improves performance its use will spread all over.

In this approach two main mechanisms for the
propagation of innovations are considered.
Firstly, external pressure can push an
agent to adopt an innovation. This mechanism intends to mimic
exogenous influence, such as advertising, and is independent of
the network structure. Secondly, there is the interaction among
agents, which depends on the underlying network structure and is
introduced in the model by considering local coupling rules. A
single tunable parameter $C$, which is fixed and the same for all
agents, accounts for the agents' resistance to change and controls
the dynamical behavior of the system. In earlier studies, some of
us have already focused on the several outcomes of the model in
the social and economic context \cite{guardiola,guardiola2,llas}.
From the economic point of view, the main result is that
the system presents an optimal behavior for an intermediate
value of $C$, and that this can be quantified with a
macroscopic observable. This feature is closely related
with the statistical properties of the profile of
technological levels of the agents and its dynamical evolution.
A proper characterization of these properties can be done
with the tools of statistical mechanics and it is the main
aim of this work.

In this paper we show that this model can be interpreted as a
surface growth model. Such interpretation
allows us to analyze the dynamical
behavior of the model as a kinetic roughening process akin to
other nonequilibrium surface growth systems. We find that the
model exhibits a continuous phase transition between a rough and a
flat phase at a critical value $C_{th}$ of the control parameter.
We focus on the scaling properties at the threshold in order to
determine the critical exponents at the transition. By defining a
convenient order parameter and studying its finite-size scaling
properties near criticality we are able to show that the
horizontal correlation length diverges as $\xi \sim \vert C -
C_{th} \vert^{-\nu}$, where $\nu \approx 2.5$. Close to the
threshold, relaxation dynamics to the stationary regime is
characterized by diverging correlation times $\tau \sim \xi^z$,
where $z \approx 0.57$ is the dynamic exponent. The existence of a
nonequilibrium roughening transition in a 1+1 dimension model
makes it interesting also for statistical mechanics. It is known
that phase transitions in nonequilibrium $1+1$ dimensional systems
are usually associated with systems with absorbing states
\cite{hinrichsen:review}. In this case, the number of absorbing
states and symmetries among them determine the universality class
to which a particular system belongs to. Thus, it is of great
interest to find models far from equilibrium which do not possess
absorbing states but still display a phase transition. As we will
see below, our model lacks absorbing states and the measured
critical exponents suggest that this model belongs to a new
universality class. Finally, we discuss the physical mechanisms
behind the critical transition in this model.

\section{The model}
We consider $N$ agents placed at the sites of a one-dimensional
lattice with periodic boundary conditions.
Each site (agent) $i$ is characterized by a real variable
$h_i$.  In general, we can consider this quantity as a {\em
characteristic} of a given individual that other agents might want
to imitate. When an agent has adopted a new feature (innovation),
her neighbors become aware of the change and balance their
interest (quantified as $h_i-h_j$) with their resistance to change
$C$ to decide if they would like to imitate this change. In this
way $C$ controls the mechanism of imitation. This parameter is
constant and the same for all the agents in the system.

The system is updated as follows \cite{guardiola2}:
\begin{enumerate}
\item At each time step an agent $h_i$ is randomly selected and
\begin{equation}
h_i \to h_i + \Delta,
\end{equation}
where $\Delta$ is a random variable uniformly distributed in
$[0,1]$\footnote{This election of the noise distribution is the
only difference with model in \cite{guardiola}. The main properties
of the model, among them the roughening transition, are robust under
this change.}.
The driving process accounts for the external pressure
that may lead an individual to spontaneously update by adopting a
new technology. This mechanism keeps the system out of
equilibrium.





\item The agents $j \in \Gamma(i)$, being $\Gamma(i)$ the set
of nearest-neighbours of agent $i$,  upgrade if $h_i - h_j \geq C$.
If the latter is satisfied, agent $j$ imitates agent $i$ by
setting $h_j = h_i$. In this way the
information of an update may spread beyond the neighbors of the
originally perturbed site. This procedure is repeated until no one
else wants to change, concluding an {\em avalanche} of imitation
events. We thus assume that the time scale of the imitation
process is much shorter than the one corresponding to the external
driving.

\end{enumerate}

\begin{figure}
\centerline{\epsfxsize=7.0cm \epsfysize=5.0cm
\epsfbox{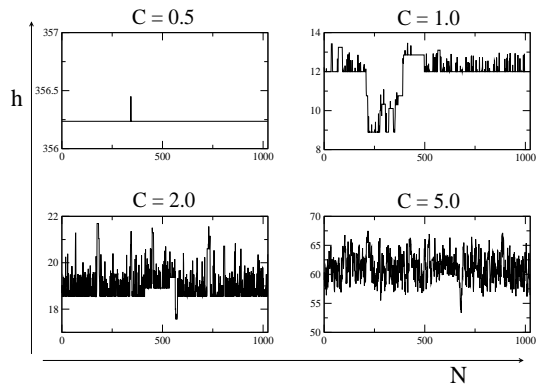}}
\caption{Snapshots of the profile of a system with $N=1024$ for
  $C=0.5$, $C=1.0$, $C=2.0$, and $C=5.0$.}
\label{fig1}
\end{figure}
Starting from a flat initial condition, $h_i=0$ for all $i$, the
system evolves to a stationary regime. In Fig. \ref{fig1} we
present snapshots of the surface profile in the stationary regime
for four different values of $C$.

The time scale separation --namely, slow driving versus fast
relaxation in the form of avalanches of activity-- is similar to
that occurring in self-organized critical (SOC) systems and
dynamically drives the system towards a stationary state
\cite{henrik}. We will see below that, at variance with most SOC
systems, two different stable phases are possible: an ordered
(flat) phase and a disordered (rough) phase with scale invariant
properties.
For small $C$, the driving process easily triggers
avalanches that cover the whole system, leading to a uniform
advance and a flat phase. On the contrary, for large $C$, there
are almost no avalanches, and the system advances mostly due to
the random updates, thus presenting an extremely heterogeneous and
rough profile. For intermediate values of C one can clearly see
the presence of large avalanches and new updates. In fact, in the
intermediate regime one can find the optimal growth regime in which
the agents reach a given average level with a minimum number of
upgrades \cite{guardiola}.

\section{Critical roughening transition}
\subsection{Stationary regime}
\label{interpolation} In order to characterize the different
morphological phases we have performed extensive numerical
simulations of the model. The fluctuations of the profile height
are measured by means of the global interface width
\cite{barabasi},
\begin{equation}
W(N,t) =  \left\langle \; \sqrt{(1/N) \sum_{i=1}^N[h_i(t) -
\overline{h}(t)]^2} \; \right\rangle
\end{equation}
Where $\langle \rangle$ stands for average over noise
realizations. At each time step the mean height value
\begin{equation}
 \overline{h}(t)= (1/N) \sum_{i=1}^N h_i(t)
\end{equation}
is also calculated. It is important to stress here that time is
always measured in the external driving temporal scale, so that
one time step $t$ corresponds to an external update.
As a consequence, the number of agents that
change their state may vary from a single one (which changes from
$h$ to $h+\Delta$) to any number of agents in the system if the
update generates an avalanche.

\begin{figure}
\epsfxsize=7.0cm \epsfysize=4.5cm \epsfbox{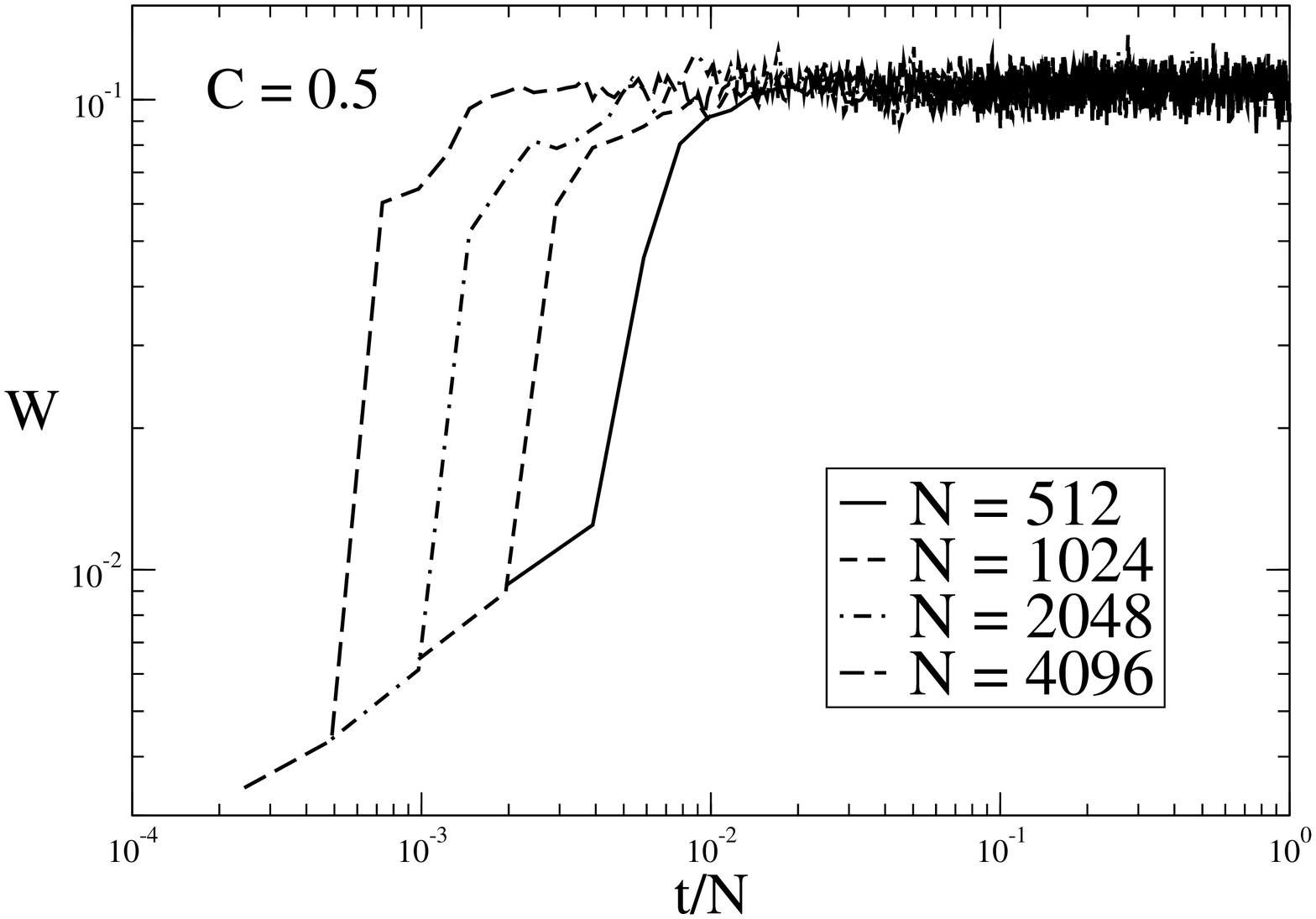}
\hspace{0.5cm}
\epsfxsize=7.0cm \epsfysize=4.5cm \epsfbox{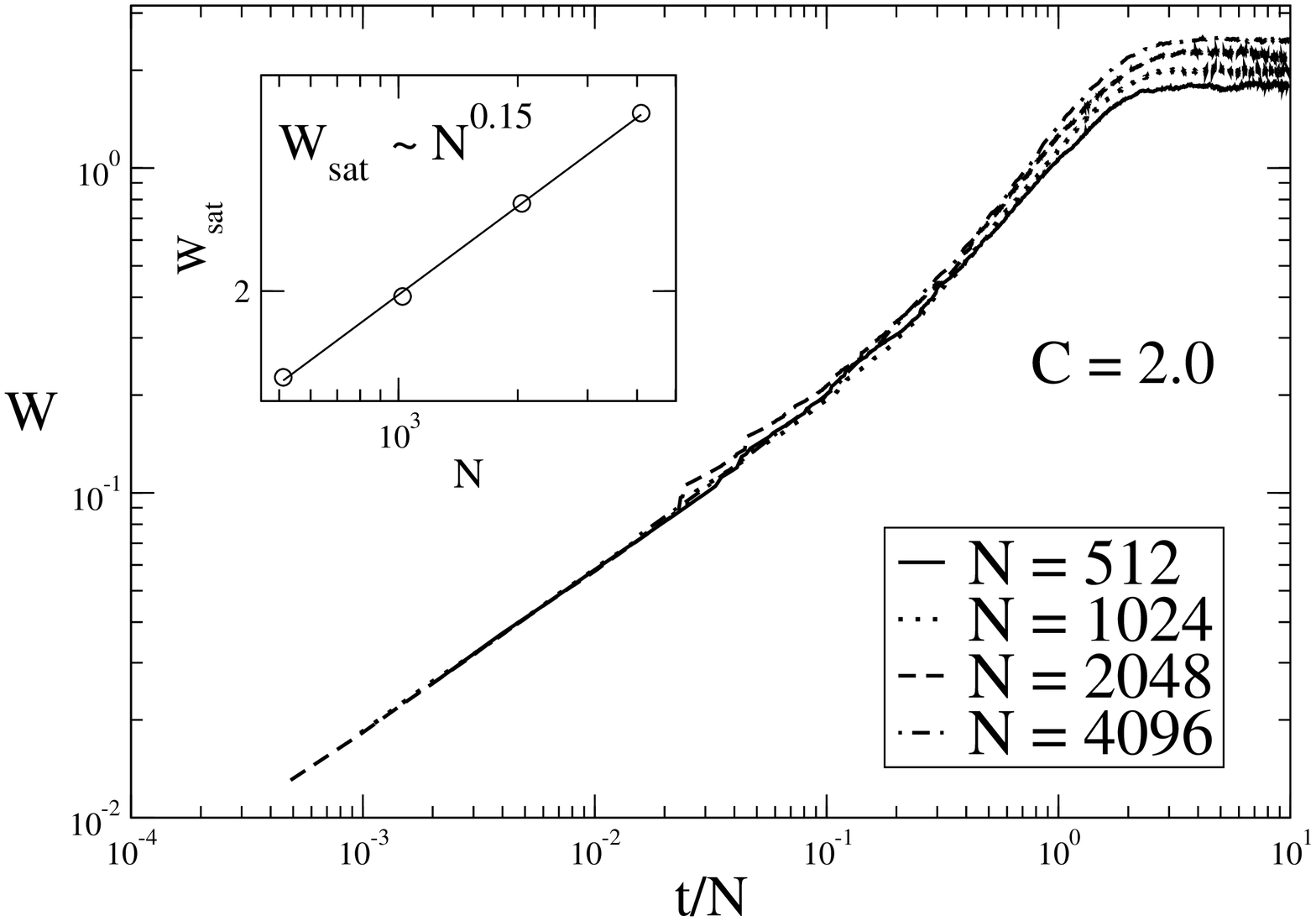}
\caption{Time evolution of the width $W(t)$  for four different
system sizes ($N = 512$, $N = 1024$, $N = 2048$, and $N = 4096$)
 when $C= 0.5$ and $C=2.0$.
Results correspond to averages over 500 realizations of the noise.
The inset shows that $W_{sat}$ scales with system size.}
\label{fig2}
\end{figure}

In the following we report on the behavior of the width in the two
different phases. 
On the top panel of Fig. \ref{fig2} we show the behavior of
$W(N,t)$  for $C=0.5$. The saturation value does not depend on the
system size, which indicates that the system is in the smooth
phase. On the bottom panel of Fig. \ref{fig2} we show the
numerical results in the rough phase, for $C=2.0$. In this case
the saturation value $W_{sat}(N)$ scales with the system size, as
is shown in the inset. We find that in the rough phase the height
fluctuations seem to fit reasonable well with a scaling as
$W_{sat}(N) \sim N^{0.15}$, which actually cannot be distinguished
from a possibly logarithmic
dependence. These results strongly suggest the presence of a
roughening transition.

In order to study the critical behavior at the transition
threshold we introduce a convenient order parameter. When looking
at the profile snapshots in the stationary regime in Fig. \ref{fig1}, one
can easily notice the presence of large plateaux, {\it i.e.}
finite connected regions of agents that have the same height. The
size of these flat regions decreases as $C$ grows, since for $C\to
\infty$ the model has to become equivalent to the random
deposition model \cite{barabasi}.
We found that the size of the largest plateau can be used as
an order parameter. In the following, we shall call
$M$ to the size of the largest plateau in the system, normalized
by the system size $N$. In this way, a completely flat profile
corresponds to $M=1$. 
We have also tried other common choices, as the often used Ising-like
magnetization (1/N)\:$\sum_i
(-1)^{h_i(t)}$ and its variations \cite{hinrichsen,juanma}.
However, we found that our election has better scaling properties
for this particular case, since it takes into account the singular
behavior of the flat phase in this model.
\begin{figure}
\centerline{\epsfxsize=7.0cm \epsfysize=5.0cm \epsfbox{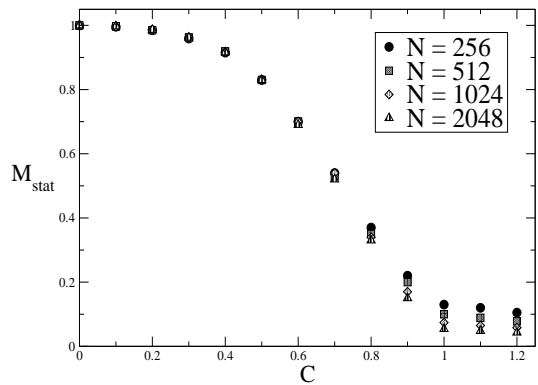}}
\caption{Order parameter $M_{stat}$ vs. $C$ for four different
system sizes, $N = 256$, $N=512$, $N = 1024$, and $N = 2048$. The
points correspond to an average over 250 realizations of the
noise.} \label{fig3}
\end{figure}

\begin{figure}
\centerline{\epsfxsize= 7.0cm \epsfysize=5.0cm \epsfbox{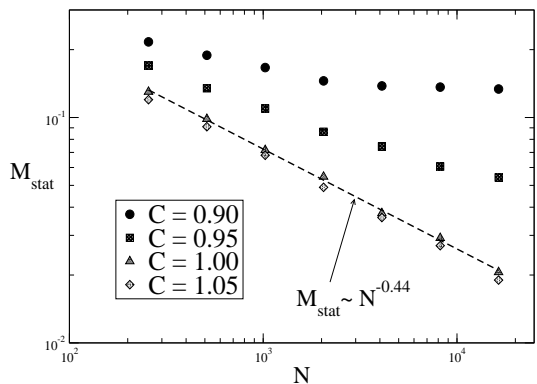}}
\caption{Order parameter $M_{stat}$ vs. $N$ for four different
values of the parameter $C=0.90$, $C=0.95$, $C=1.00$ and $C=1.05$.
A power law decay $M_{stat}\sim N^{-0.44}$ is observed for
$C=1.0$. Results correspond to averages over 500 realizations.}
\label{fig4}
\end{figure}

\begin{figure}
\centerline{\epsfxsize= 7.0cm \epsfysize=5.0cm \epsfbox{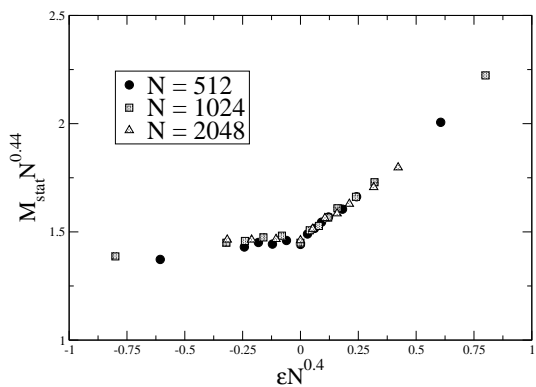}}
\caption{Data collapse of the order parameter $M_{stat}$ as
given in Eq. (\ref{scaling1}). Results correspond to averages over
500 realizations.}
\label{fig5}
\end{figure}

Starting from a flat initial condition $M(t)$ evolves until it
reaches a stationary value. In Fig. \ref{fig3} we show the
behavior of the stationary value of the order parameter
$M_{stat}(N,C)$ vs. $C$ for four different system sizes. The order
parameter allows us to distinguish the two phases discussed above.
Note that for small values of the control parameter, the system
gets ordered, implying a flat phase. On the contrary, the
stationary value of the order parameter goes to zero for large
values of $C$ as the system size becomes larger. Critical behavior
is expected close to the threshold $C_{th}$ and, as usual, it can
be studied numerically by finite-size scaling techniques
\cite{png,hinrichsen,juanma} as follows. For any value of the
control parameter $C$, there exists a horizontal correlation
length $\xi$, which diverges as $\xi \sim \epsilon^{-\nu}$ when
the distance to the critical threshold goes to zero $\epsilon =
\vert C - C_{th} \vert \to 0$. In finite systems this actually
occurs for values of $C$ close to, but not exactly at, the
threshold since the finite-size critical behavior is encountered
as long as $\xi \sim N$, or equivalently when $\epsilon \sim
N^{-1/\nu}$. Close to the threshold, $\epsilon \to 0$,
for sufficiently large values of the system size,
$M_{stat}$ converges to a finite value obeying
$M_{stat}(N,\epsilon) \sim \epsilon^{\beta}$.
Just at the critical point $\epsilon = 0$ we expect
the order parameter to decay as a power-law with the system size
\begin{equation}
M_{stat}(N,\epsilon=0) \sim N^{-\beta / \nu}.
\end{equation}
In Fig. \ref{fig4} we plot our numerical results for
$M_{stat}(N,\epsilon)$ vs. $N$ for different values of the
distance to the threshold $\epsilon$. Only for $C=C_{th}$ a
power-law with the system size can be obtained and the slope of
the straight line in a log-log plot gives an estimation of the
ratio $\beta / \nu=0.44 \pm 0.05$ between critical exponents. We
can thus identify $C_{th}=1.0 \pm 0.1$ with the critical point.
After having determined the critical point, numerical data for
different system sizes can be cast in the finite-size scaling
ansatz
\begin{equation}
M_{stat}(N,\epsilon)= N^{-\beta/\nu}g\left(\epsilon \:
N^{1/\nu}\right), \label{scaling1}
\end{equation}
where the scaling function $g(y) \sim {\rm const}$ for $y \ll 1$,
and $g(y) \sim y^{\beta}$ if $y \gg 1$.  In Fig. \ref{fig5} we
plot a data collapse that allows us to determine the values of the
exponents $1/\nu = 0.40 \pm 0.05$ and $\beta / \nu = 0.44 \pm
0.05$. From these, we then have $\beta \sim 1.10$ and $\nu \sim
2.50$.

\subsection{Dynamics}
\begin{figure}
\centerline{\epsfxsize= 7.0cm \epsfysize=5.0cm \epsfbox{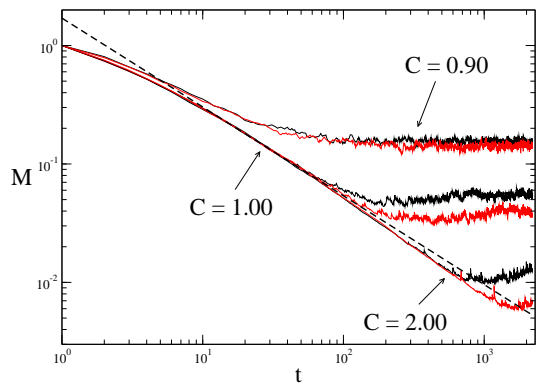}}
\caption{Order parameter dynamics for three values of $C$:
$C=0.90$ in the smooth phase, the critical value $C=1.00$ and
$C=2.00$ in the rough phase. Two system sizes, $N=2048$ and
$N=4096$, are represented to better appreciate the deviations from
power-law behavior, indicated with a dashed line,  for the values outside
the critical region. The curves correspond to an average over 500
realizations.}
\label{fig6}
\end{figure}

\begin{figure}
\centerline{\epsfxsize= 7.0cm \epsfysize=5.0cm \epsfbox{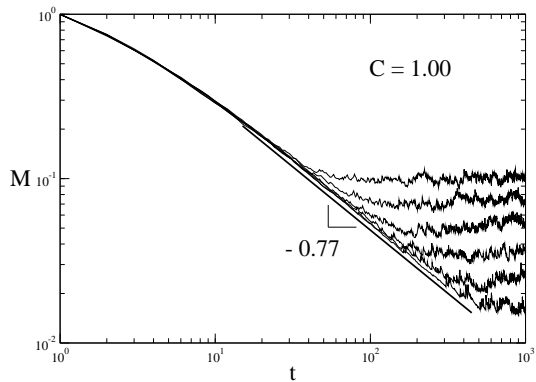}}
\caption{Order parameter dynamics for six increasing values of the
system size $N$, from top to bottom, $2^9$ to $2^{14}$. A power-law
behavior $M\sim t^{-0.77}$ is observed in the transient regime.
Results correspond to averages over 500 realizations.}
\label{fig7}
\end{figure}

\begin{figure}
\centerline{\epsfxsize= 7.0cm \epsfysize=5.0cm \epsfbox{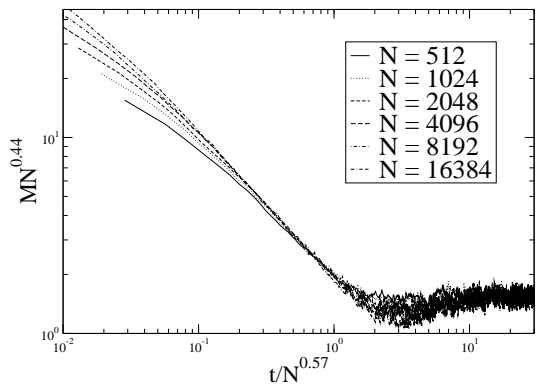}}
\caption{Dynamical data collapse of the order parameter
at the critical point, as given by Eq. (\ref{scaling2}). The
exponents used correspond to the ones obtained above,
$\beta/\nu \sim 0.44$ and $z \sim 0.57$.}
\label{fig8}
\end{figure}

Since the model is out of equilibrium our study is completed next
with an analysis of the dynamic behavior, which contains much
information about the universality of the roughening transition.
In Fig. \ref{fig6} we show the temporal behavior of the order
parameter $M(t,N,C)$ for three different values of $C$ (above,
below and at the critical threshold). Again, only at the critical
point we may expect to find a power-law decay $M(t,N,\epsilon=0)
\sim t^{- \beta / \nu_{t}}$, where $\nu_{t}$ is the exponent
associated with the diverging correlation time $\tau \sim
\epsilon^{-\nu_{t}}$ as $\epsilon \to 0$. The correlation time
corresponds to the typical time that correlations survive in the
system and is given by $\tau \sim \xi^z$, where $z$ is the dynamic
exponent. The three exponents are related by the usual scaling
relation $z = \nu_{t} / \nu$ provided dynamic scaling holds. In
Fig. \ref{fig6} we can see that only at the critical point a
power-law behavior is observed, while deviations occur for $C \neq
C_{th}$. The fit to a straight line in a log-log plot, as shown
Fig. \ref{fig7}, leads to a
determination of the ratio $\beta / \nu_t = 0.77 \pm 0.05$. One can
write the dynamic scaling ansatz
\begin{equation}
M(t,N,\epsilon) = N^{-\beta / \nu} \Phi \left( \epsilon t^{1 /
\nu_t}, t / N^z \right)
\end{equation}
for the order parameter, which at the critical point, $\epsilon =
0$, reads
\begin{equation}
M(t,N,\epsilon=0) = N^{-\beta / \nu} f \left(t / N^z \right),
\label{scaling2}
\end{equation}
where the scaling function $f(u) \sim {\rm const}$ for $u \gg 1$
and $f(u) \sim u^{- \beta / \nu_t}$ for $u \ll 1$. We can then use
the values of the exponents just obtained to collapse our data as
shown in Fig. \ref{fig8} with exponents $\nu_t \sim  1.43$ and $z \sim
0.57$. 

Nonequilibrium phase transitions have been mostly related to the
universality class of directed percolation (DP), with very few
exceptions \cite{hinrichsen:review}. In particular, there are many
examples of roughening transitions far from equilibrium that have
been linked to DP, examples include polynuclear growth models
\cite{png}, solid-on-solid models with evaporation at the edges of
terraces \cite{hinrichsen} and the fungal growth model
\cite{juanma}. In all these systems, the DP process emerges at a
particular reference height of the interface. In this case, the
critical exponents characterizing the roughening transition can be
obtained from those of DP, which in 1+1 dimension are given by
$\nu = \nu_{\perp}^{DP} = 1.10$ for the correlation length
exponent, $\nu_{t} = \nu_{\parallel}^{DP} = 1.73$ for the time
correlation exponent, and $z = z^{DP} = 1.58$ the dynamic exponent
\cite{hinrichsen:review}. Our results clearly suggest that the
roughening transition occurring in the innovation propagation
model does not belong to the DP class.

The relation of many nonequilibrium critical models to DP has led
to the proposal of the conjecture due to Janssen and Grassberger
\cite{janssen,grassberger}, which states that a model belongs to
DP under the following assumptions \cite{hinrichsen:review}:
\begin{enumerate}
\item The model displays a continuous phase transition from a
fluctuating active phase into a unique absorbing state. \item The
transition is characterized by a positive one-component order
parameter. \item The dynamic rules involve only short-range
interactions. \item Finally, the system has no special attributes
like additional symmetries or quenched randomness.
\end{enumerate}
Any model satisfying all above four conditions has been found to
belong to DP universality class, with no exception to date.
However, it is known that at least some of the above DP conditions
can be relaxed. In fact, there are a few examples of systems that,
despite exhibiting no absorbing states
\cite{png,hinrichsen,juanma} or having quenched disorder
\cite{tang,buldyrev} also display nonequilibrium phase transitions
that belong to the DP universality class. Our model does not have
absorbing states, since in both the rough and the flat phase the
interface keeps fluctuating. Also and perhaps most importantly,
interaction is not short-ranged, because of the avalanches of
activity that give rise to nonlocal effects with finite
probability.
Their influence in the dynamics is reflected by the extremly low
value of the dynamic exponent, $z = 0.57$ $<2$, signature of a
highly super-diffusive behavior. It appears that this nonlocal
interaction mechanism is the responsible for the deviation of
the DP critical behavior.

\begin{figure}
\centerline{\epsfxsize= 7.0cm \epsfysize=5.0cm \epsfbox{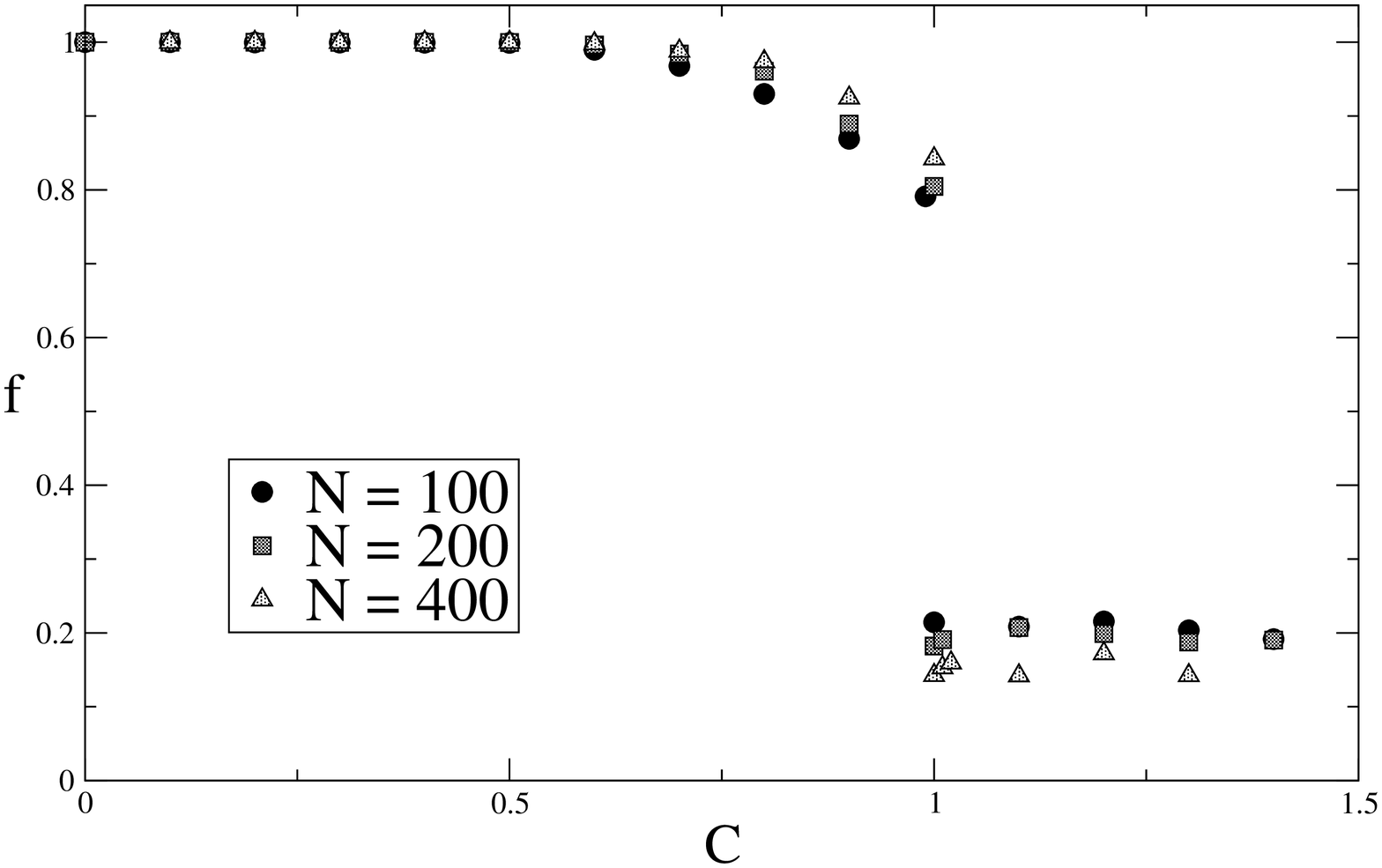}}
\caption{Fraction of sites $f$ that can generate an avalanche with
a single update as a function of $C$ for three different system
sizes, $N = 100$, $N = 200$ and $N = 400$. Curves are averaged
over $1000$ realizations.}
\label{fig9}
\end{figure}

We believe that the transition takes place exactly at $C=1.0$.
This is directly related to the dynamical evolution rules of the
model. We have defined the external driving by choosing a random
number from a uniform distribution in $[0,1]$. As a consequence,
for $C<1.0$, a random update on any site can generate an
avalanche. On the other hand, for $C>1.0$, only a small fraction
of sites will be able to generate an avalanche with a single
update. In order to quantify this effect we have studied the
fraction of sites which can generate an avalanche with a single
update. A site $i$ with this property will satisfy
\begin{equation}
  h_i - h_{i\pm1} + 1 > C
\end{equation}
In Fig. \ref{fig9} we present the fraction of sites $f$ which are
able to generate an avalanche as a function of C. The figure clearly shows
that this fraction remains close to one for $C < 1.0$ and drops abruptly
to a small value for $C > 1.0$.

\section{Conclusions}
In this work we have studied a simple model of innovation
propagation dynamics in an economic system as a surface growth
model. This has allowed us to characterize different morphological
phases and also to analyze the dynamical behavior of the model as
a kinetic roughening process. We have characterized a roughening
transition and determined its critical exponents by finite-size
scaling techniques. The value of the exponents do not coincide
with known universality classes. We believe that the avalanche
driven dynamics with its long-range effects is the reason why this
model does not belong to the DP universality class. We have also
presented a possible mechanism for the transition occurring
exactly at $C_{th}=1$.

\acknowledgements We would like to thank X.\ Guardiola, J.\ J.\
Ramasco, M.\ A.\ Rodr\'{\i}guez, A.\ Arenas, C.\ J.\ P\'erez, and
F.\ Vega-Redondo for fruitful discussions at the earliest
stages of this work.
M.\ Llas, P.\ M.\ Gleiser, and A.\ D\'{\i}az-Guilera
acknowledge financial support from MCYT, grant number BFM2000-0626,
and also from European Commission, Fet Open Project COSIN IST-2001-33555.
P.\ M.\ G.\ acknowledges financial support from Fundaci\'on Antorchas.
J.\ M.\ L.\ acknowledges support from the Ministerio de
Ciencia y Tecnolog{\'\i}a (Spain) and FEDER under project
BFM2000-0628-C03-02.

\end{document}